\documentclass[preprint,prd,nofootinbib,tightenlines,amsmath]{revtex4}
\usepackage{axodraw}
\usepackage{graphics}
\usepackage{epsfig}
\usepackage{dcolumn}
\usepackage{bm}

\begin{document}
\baselineskip=15pt \parskip=5pt

\hspace*{\fill} $\hphantom{-}$

\def\lsim{\mathrel {\vcenter {\baselineskip 0pt \kern 0pt
    \hbox{$<$} \kern 0pt \hbox{$\sim$} }}}
\def\gsim{\mathrel {\vcenter {\baselineskip 0pt \kern 0pt
    \hbox{$>$} \kern 0pt \hbox{$\sim$} }}}

\preprint{hep-ph/yymmnnn}

\title{The Decay  $\bm{\Omega^-\to\Xi^-\pi^+\pi^-}$ in Chiral Perturbation Theory}

\author{Oleg Antipin}
\email{oaanti02@iastate.edu}
\affiliation{Department of Physics and Astronomy, Iowa State University, Ames, IA 50011, USA}

\author{Jusak Tandean}
\email{jtandean@ulv.edu}
\affiliation{Department of Mathematics/Physics/Computer Science, University of La Verne,
La Verne, CA 91750, USA}

\author{G. Valencia}
\email{valencia@iastate.edu}
\affiliation{Department of Physics and Astronomy, Iowa State University, Ames, IA 50011,~USA}

\author{$ $}

\date{\today}

\begin{abstract}

We study the decay  $\Omega^-\to\Xi^-\pi^+\pi^-$ in heavy-baryon chiral perturbation theory.
At leading order, the decay is completely dominated by the  $\Xi^{*0}(1530)$ intermediate
state, and the predicted rate and $\Xi^-\pi^+$-mass distribution are in conflict with
currently available data. It is possible to resolve this conflict by considering additional
contributions at next-to-leading order.

\end{abstract}


\maketitle

\section{Introduction}

It was suggested many years ago that the decay  \,$\Omega^-\to\Xi^-\pi^+\pi^-$\,
should be dominated by the $\Xi^{*0}(1530)$ intermediate
state~\cite{Goswami:1972tg,Finjord:1979pz}.
Under this assumption, the current Particle Data Group~\cite{pdg} branching ratio for
\,$\Omega^-\to\Xi^{*0}\pi^-$\,  has been deduced from the measurement of
${\cal B}(\Omega^- \to \Xi^- \pi^+ \pi^-)$~\cite{Bourquin:1984gd}.
More recently, the HyperCP collaboration has reported a preliminary measurement of
\,$\Omega^-\to\Xi^-\pi^+\pi^-$\, that is very surprising in that the distribution of
the $\Xi^-\pi^+$ invariant-mass apparently shows no evidence for
the $\Xi^{*0}(1530)$ dominance~\cite{hypercp}.

Motivated by this result, we revisit the calculation of the rate for this decay mode
using heavy-baryon chiral perturbation theory (HB$\chi$PT).
We first present a leading-order calculation that reproduces the expectation that the decay is
completely dominated by the $\Xi^{*0}(1530)$ intermediate state.

We next explore whether higher-order contributions can reconcile the calculation with
the preliminary HyperCP result. To this end, we consider the effect of next-to-leading-order
diagrams, which occur at tree level.

\section{Leading-order calculation\label{loc}}

The amplitude for
\,\,$\Omega^-\to\Xi^-(p_\Xi^{}\bigr)\,\pi^+(p_+^{}\bigr)\,\pi^-(p_-^{}\bigr)$ can be written
in the heavy-baryon approach as
\begin{eqnarray}  \label{M_O}
{\cal M}(\Omega^-\to\Xi^-\pi^+\pi^-)  \,\,=\,\,
-\bar{u}_\Xi^{}\, \bigl( A_+^{}\, p_+^\mu + A_-^{}\, p_-^\mu
+ 2B_+^{}\, S_v^{}\cdot p_-^{}\, p_+^\mu
+ 2B_-^{}\, S_v^{}\cdot p_+^{}\, p_-^\mu \bigr)\, u_{\Omega\mu}^{}   \,\,,
\end{eqnarray}
where  $A_\pm$ and $B_\pm$ are independent form-factors and $S_v$  is the spin operator.
The most general form of the amplitude has eight independent form-factors~\cite{Goswami:1972tg},
and we have included here only the ones that receive contributions from the leading-order
and next-to-leading-order diagrams that we consider.
The partial decay width resulting from the amplitude above is
\begin{eqnarray}
d\Gamma(\Omega^-\to\Xi^-\pi^+\pi^-)  \,\,=\,\,
\frac{1}{32\, \bigl(2\pi\, m_\Omega^{} \bigr)^3}\,
\overline{|{\cal M}(\Omega^-\to\Xi^-\pi^+\pi^-)|^2}\, dm_{\Xi^-\pi^+}^2\,dm_{\Xi^-\pi^-}^2 \,\,,
\end{eqnarray}
where  \,$m_{\Xi^-\pi^\pm}^2=\bigl(p_\Xi^{}+p_\pm^{}\bigr)^2$\,  and
\begin{eqnarray}
\overline{|{\cal M}(\Omega^-\to\Xi^-\pi^+\pi^-)|^2}  &=&
\mbox{$\frac{4}{3}$}\, m_\Omega^{}\,m_\Xi^{}\, \Bigl\{ \bigl|A_+^{}\bigr|^2\, \bm{p}_+^2
+ \bigl|A_-^{}\bigr|^2\, \bm{p}_-^2
+ 2\, {\rm Re}\bigl( A_+^* A_-^{} \bigr)\, \bm{p}_+^{}\!\cdot\!\bm{p}_-^{}
\nonumber \\ && \hspace*{5em}
+\,\,
\Bigl[ \bigl|B_+^{}\bigr|^2+\bigl|B_-^{}\bigr|^2+{\rm Re}\bigl( B_+^*B_-^{} \bigr) \Bigr]\,
 \bm{p}_+^2\, \bm{p}_-^2
\nonumber \\ && \hspace*{5em}
+\,\, {\rm Re}\bigl(B_+^*B_-^{}\bigr)\, \bigl(\bm{p}_+^{}\cdot\bm{p}_-^{}\bigr)^2\Bigr\} \,\,,
\end{eqnarray}
with  $\bm{p}_\pm^{}$  denoting the three-momenta of the pions in the $\Omega^-$ rest frame.

The chiral Lagrangian describing the interactions of the lowest-lying mesons and baryons
is written down in terms of the lightest meson-octet, baryon-octet, and baryon-decuplet
fields~\cite{Gasser:1983yg,Bijnens:1985kj,Jenkins:1991ne}.
The meson and baryon octets are collected into  $3\times3$  matrices $\varphi$ and $B$,
respectively, and the decuplet fields are represented by the Rarita-Schwinger tensor
$T_{abc}^\mu$, which is completely symmetric in its SU(3) indices ($a,b,c$).
The octet mesons enter through the exponential $\,\Sigma=\xi^2=\exp(i\varphi/f),\,$
where  $\,f=f_\pi^{}=92.4\rm\,MeV\,$ is the pion-decay constant.

In the heavy-baryon formalism~\cite{Jenkins:1991ne}, the baryons in
the chiral Lagrangian are described by velocity-dependent fields, $B_v^{}$  and  $T_v^\mu$.
For the strong interactions, the Lagrangian at lowest order in
the derivative and  $m_s^{}$  expansions is given by
\begin{eqnarray}   \label{Ls}
{\cal L}_{\rm s}^{}  &=& \left\langle \bar B_v^{}\, i v^\mu \bigl(
\partial_\mu^{}B_v^{}+\bigl[{\cal V}_\mu^{},B_v^{} \bigr] \bigr) \right\rangle
+ 2D \left\langle \bar B_v^{} S_v^\mu \left\{ {\cal A}_\mu^{}, B_v^{} \right\} \right\rangle
+ 2F \left\langle \bar B_v^{} S_v^\mu \left[ {\cal A}_\mu^{}, B_v^{} \right] \right\rangle
\nonumber \\ && -\,\,
\bar T_v^\mu\,  i v\cdot{\cal D}T_{v\mu}^{} +
\Delta m\, \bar T_v^\mu T_{v\mu}^{} +
{\cal C} \left( \bar T_v^\mu {\cal A}_\mu^{} B_v^{}
               + \bar B_v^{} {\cal A}_\mu^{} T_v^\mu \right)
+ 2{\cal H}\; \bar T_v^\mu S_v^{}\cdot{\cal A}T_{v\mu}^{}
\nonumber \\ && +\,\,
b_D^{} \left\langle \bar B_v^{} \left\{ M_+^{}, B_v^{} \right\}\right\rangle
+ b_F^{} \left\langle \bar B_v^{} \left[ M_+^{}, B_v^{} \right] \right\rangle
\,+\,  c\, \bar T_v^\mu M_+^{} T_{v\mu}^{}
\end{eqnarray}
where only the relevant terms are shown,
$\,\langle\cdots\rangle\equiv{\rm Tr}(\cdots)\,$  in flavor-SU(3) space,
$\Delta m$  denotes the mass difference between the decuplet
and octet baryons in the chiral limit,
$\,{\cal V}^\mu =
\frac{1}{2}\bigl(\xi\,\partial^\mu\xi^\dagger+\xi^\dagger\,\partial^\mu\xi\bigr),\,$
$\,{\cal A}^\mu =
\frac{i}{2}\bigl(\xi\,\partial^\mu\xi^\dagger-\xi^\dagger\,\partial^\mu\xi\bigr),\,$
$\,{\cal D}^\mu T_{klm}^\nu =
\partial^\mu T_{klm}^\nu + {\cal V}_{kn}^\mu T_{lmn}^\nu
+ {\cal V}_{ln}^\mu T_{kmn}^\nu + {\cal V}_{mn}^\mu T_{kln}^\nu,\,$
and $\,M_+ = \xi^\dagger M\xi^\dagger + \xi M^\dagger\xi,\,$  with
$\,M={\rm diag}(\hat{m},\hat{m},m_s^{})=
{\rm diag}\bigl(m_\pi^2,m_\pi^2,2m_K^2-m_\pi^2\bigr)/\bigl(2B_0\bigr)\,$
in the isospin-symmetric limit  $\,m_u^{}=m_d^{}=\hat{m}$.\,\,
The constants  $D$, $F$, $\cal C$, $\cal H$, $B_0$, $b_{D,F}$, and $c$ are free
parameters which can be extracted from data.

As is well known, the weak interactions responsible for hyperon nonleptonic decays are
described by a  $\,|\Delta S|=1\,$ Hamiltonian that transforms as
$(8_{\rm L},1_{\rm R})\oplus(27_{\rm L},1_{\rm R})$
under  SU(3$)_{\rm L}$$\times$SU(3$)_{\rm R}$  rotations.
It is also known empirically that the octet term dominates the 27-plet term.
We therefore assume in what follows  that the decays are completely characterized by
the  $(8_{\rm L}^{},1_{\rm R}^{})$, $\,|\Delta I|=1/2\,$  interactions.
The leading-order chiral Lagrangian for such interactions
is~\cite{Bijnens:1985kj,Jenkins:1991bt}
\begin{eqnarray}  \label{weakcl}
{\cal L}_{\rm w}^{}  &=&
h_D^{} \left\langle \bar B_v^{} \left\{
\xi^\dagger h \xi\,,\,B_v^{} \right\} \right\rangle
+ h_F^{} \left\langle \bar B_v^{} \left[
\xi^\dagger h \xi\,,\,B_v^{} \right] \right\rangle
+ h_C^{}\, \bar T_v^\mu\, \xi^\dagger h \xi\, T_{v\mu}^{}
\;+\;  {\rm H.c.}   \;,
\end{eqnarray}
where  $h$  is a  3$\times$3  matrix having elements  $\,h_{kl}=\delta_{k2}\delta_{3l}$\,
and the parameters  $h_{D,F,C}$ can be fixed from two-body hyperon nonleptonic decays.

\begin{figure}[tb]
\begin{picture}(170,60)(-30,-15) \SetWidth{1}
\Text(-10,0)[]{\small$\Omega^-$} \Line(0,0)(100,0)
\DashLine(30,0)(30,25){2} \Text(33,32)[]{\small$\pi^-$}
\DashLine(70,0)(70,25){2} \Text(73,32)[]{\small$\pi^+$} \Text(110,0)[]{\small$\Xi^-$}
\Text(50,-6)[]{\small$\Xi^{*0}$} \Vertex(70,0){2} \SetWidth{1} \BBoxc(30,0)(4,4)
\end{picture}
\\
\begin{picture}(220,60)(-40,-15) \SetWidth{1}
\Text(-10,0)[]{\small$\Omega^-$} \Line(0,0)(140,0)
\DashLine(70,0)(70,25){2} \Text(73,32)[]{\small$\pi^-$}
\DashLine(110,0)(110,25){2} \Text(113,32)[]{\small$\pi^+$}
\Text(90,-6)[]{\small$\Xi^{*0}$} \Text(50,-6)[]{\small$\Xi^{*-}$} \Text(150,0)[]{\small$\Xi^-$}
\Vertex(70,0){2} \Vertex(110,0){2} \SetWidth{1} \BBoxc(30,0)(4,4)
\end{picture}
\begin{picture}(220,60)(-40,-15) \SetWidth{1}
\Text(-10,0)[]{\small$\Omega^-$} \Line(0,0)(140,0)
\DashLine(70,0)(70,25){2} \Text(73,32)[]{\small$\pi^-$}
\DashLine(110,0)(110,25){2} \Text(113,32)[]{\small$\pi^+$}
\Text(90,-6)[]{\small$\Xi^0$} \Text(50,-6)[]{\small$\Xi^{*-}$} \Text(150,0)[]{\small$\Xi^-$}
\Vertex(70,0){2} \Vertex(110,0){2} \SetWidth{1} \BBoxc(30,0)(4,4)
\end{picture}
\caption{Diagrams contributing to  \,$\Omega^-\to\Xi^-\pi^+\pi^-$\,  at leading order in
$\chi$PT.
Each solid dot represents a strong vertex from  ${\cal L}_{\rm s}$  in  Eq.~(\ref{Ls}), and
each square a weak vertex from  ${\cal L}_{\rm w}$  in  Eq.~(\ref{weakcl}).\label{lodiags}
}
\end{figure}
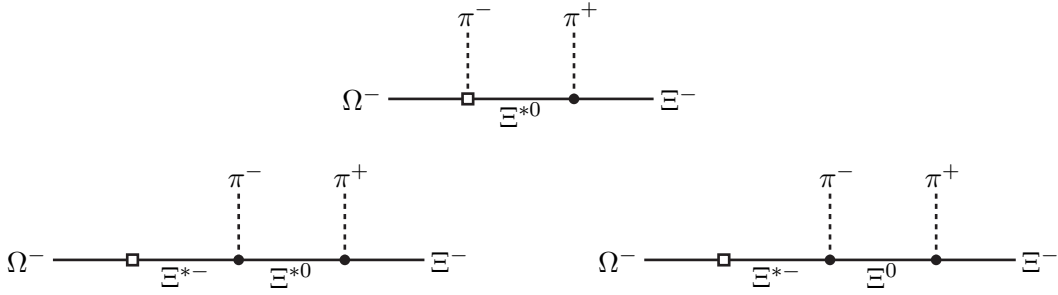

From  ${\cal L}_{\rm w}$  together with  ${\cal L}_{\rm s}$, we can derive
the ${\cal O}(p^0)$ diagrams displayed in Fig.~\ref{lodiags}.
They provide the leading-order contributions to the $A_\pm$ and $B_\pm$ form factors in
Eq.~(\ref{M_O}), namely
\begin{subequations} \label{loamps}
\begin{eqnarray}
A_+^{(0)}  \,\,=\,\,
\frac{+{\cal C}\, h_C^{}}{6\,f^2\,\bigl(E_\Xi^{}+E_+^{}-\bar{m}_{\Xi^*}^{}\bigr)} \,\,,
\end{eqnarray}
\begin{eqnarray}
A_-^{(0)}  \,\,=\,\, 0  \,\,,
\end{eqnarray}
\begin{eqnarray}
B_+^{(0)}  \,\,=\,\,
\frac{-{\cal C}\, {\cal H}\, h_C^{}}
{18\,f^2\,\bigl(m_\Omega^{}-m_{\Xi^*}^{}\bigr)\bigl(E_\Xi^{}+E_+^{}-\bar{m}_{\Xi^*}^{}\bigr)} \,\,,
\end{eqnarray}
\begin{eqnarray}
B_-^{(0)}  &=&
\frac{-{\cal C}\, (D-F)\, h_C^{}}
     {6\,f^2\,\bigl(m_\Omega^{}-m_{\Xi^*}^{}\bigr)\bigl(E_\Xi^{}+E_+^{}-m_\Xi^{}\bigr)}\,
\nonumber \\ &&+\,\,
\frac{{\cal C}\, {\cal H}\, h_C^{}}
{27\,f^2\,\bigl(m_\Omega^{}-m_{\Xi^*}^{}\bigr)\bigl(E_\Xi^{}+E_+^{}-\bar{m}_{\Xi^*}^{}\bigr)} \,\,,
\end{eqnarray}
\end{subequations}
where $\bar{m}_{\Xi^*}={m}_{\Xi^*}-\frac{i}{2}\Gamma_{\Xi^*}$.

Numerically, to evaluate the decay rates resulting from the form factors above, we employ
the tree-level values of the strong and weak parameters.
Specifically,
\begin{eqnarray}   \label{DFCtree}
D \,\,=\,\, 0.80 \,\,, \hspace{2em} F \,\,=\,\, 0.46 \,\,, \hspace{2em}
|{\cal C}| \,\,=\,\, 1.7
\end{eqnarray}
from hyperon semileptonic decays and the strong decays  $\,T\to B\varphi,\,$
but a tree-level value of $\cal H$ is not yet available from data.
Since nonrelativistic quark models~\cite{Jenkins:1991ne} give \,$3F=2D$,\,  ${\cal C}=-2D$,\,
and  \,${\cal H}=-3D$,\, which are well satisfied by  $D$, $F$, and  $\cal C$, we adopt
\begin{eqnarray}   \label{Htree}
{\cal H}  \,\,=\,\,  -2.4  \,\,.
\end{eqnarray}
For the weak parameters, we have
\begin{eqnarray} \label{weakt}
h_C^{}\,\,=\,\, 3.42\times10^{-8}{\rm~GeV}  \,\,,
\end{eqnarray}
$h_D=-1.45\times10^{-8}$\,GeV,\, and  \,$h_F=3.50\times10^{-8}$\,GeV,\,  extracted from
a simultaneous tree-level fit to the $S$-wave octet-hyperon and $P$-wave $\Omega^-$
nonleptonic two-body decays, as $h_{D,F}$ contribute not only to the octet-hyperon decays,
but also to  \,$\Omega^-\to\Lambda\bar K$,\, whereas  $h_C$  contributes to
\,$\Omega^-\to\Lambda\bar K,\Xi\pi$\,~\cite{Jenkins:1991bt}.
As seen above, $h_C$ is the only weak parameter in the lowest-order contributions to
\,$\Omega^-\to\Xi^-\pi^+\pi^-$.

\begin{figure}[tb]
\begin{picture}(1,1)(0,0) \Text(0,190)[]{\footnotesize(a)} \end{picture}
\includegraphics[width=4.5in]{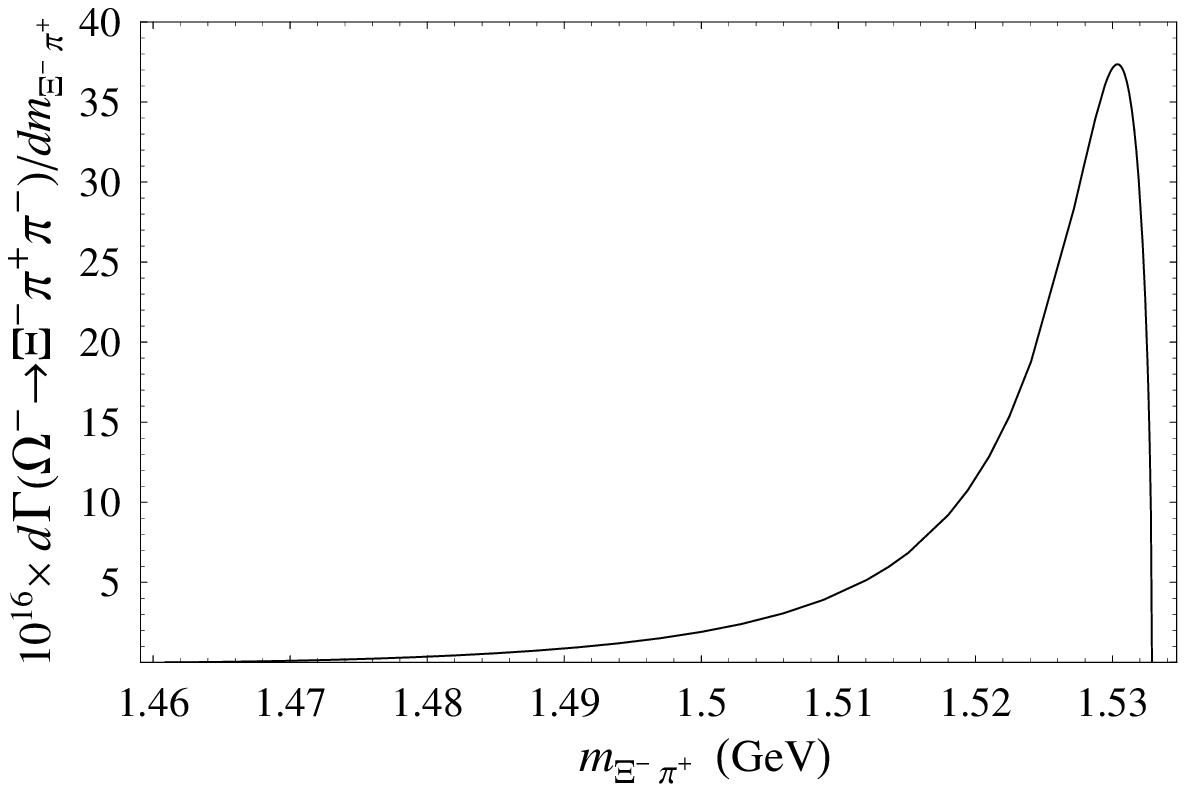} \\ \medskip
\begin{picture}(1,1)(0,0) \Text(0,190)[]{\footnotesize(b)} \end{picture}
\includegraphics[width=4.5in]{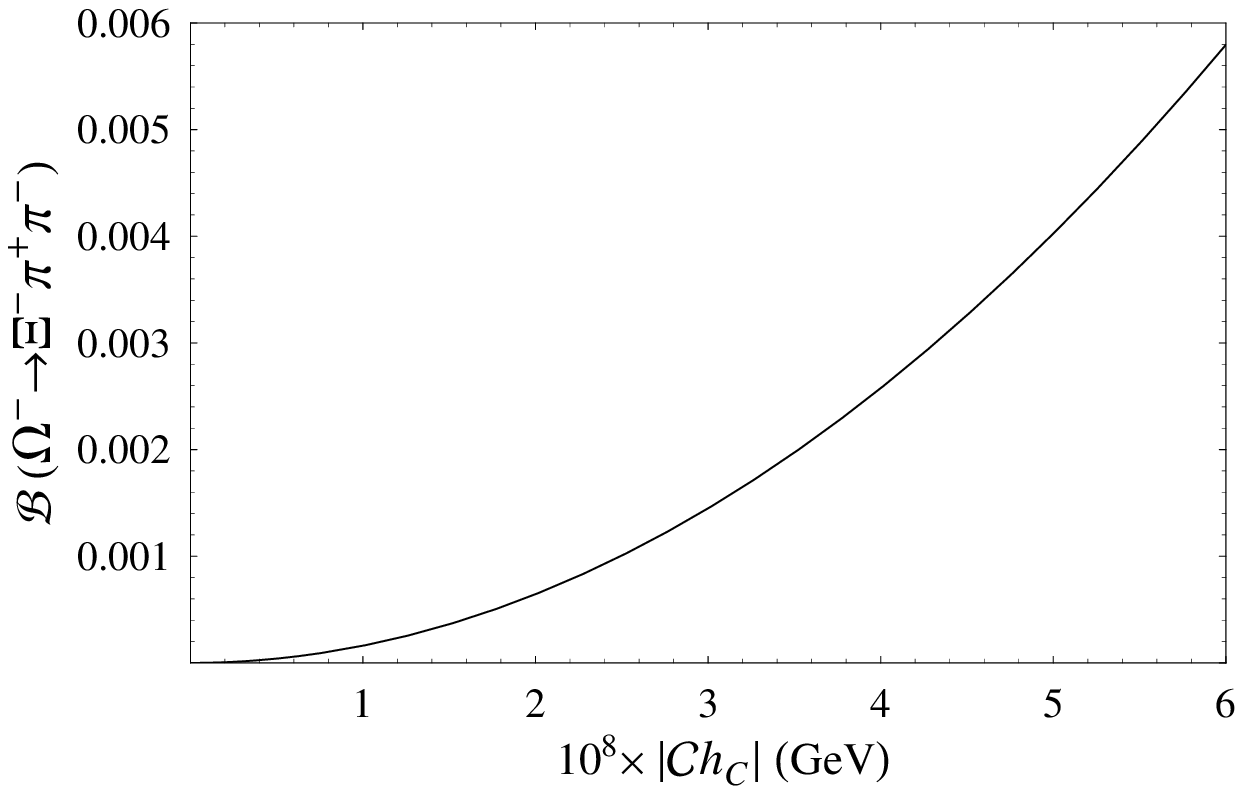}
\caption{\label{lodiff}(a) Distribution of $\Xi^-\pi^+$ invariant-mass in
\,$\Omega^-\to\Xi^-\pi^+\pi^-$\,  at leading order with parameter values in
Eqs.~(\ref{DFCtree})-(\ref{Htree}), and
(b)~its branching ratio as function of  \,$|{\cal C}h_C|$\,  with
$D$$-$$F$ and $\cal H$ values in Eqs.~(\ref{DFCtree}) and~(\ref{Htree}).}
\end{figure}

The resulting branching ratio,
\begin{eqnarray}
{\cal B}(\Omega^-\to\Xi^-\pi^+\pi^-)  \,\,=\,\,  5.4 \times 10^{-3}  \,\,,
\end{eqnarray}
is roughly an order of magnitude larger than the preliminary number reported by HyperCP,
\,${\cal B}(\Omega^-\to\Xi^-\pi^+\pi^-)=[3.6\pm0.3({\rm stat})]\times10^{-4}$\,~\cite{hypercp},
and also the current PDG value,
\,${\cal B}(\Omega^-\to\Xi^-\pi^+\pi^-)=
\bigl(4.3_{-1.3}^{+3.4}\bigr)\times10^{-4}$\,~\cite{pdg}.
In Fig.~\ref{lodiff}(a), we display the corresponding $\Xi^-\pi^+$ invariant-mass distribution.
As expected, these results are dominated by the $\Xi^*$ resonance.
Notice that the leading-order rate is proportional to $|{\cal C} h_C|^2$ so that there is
a large parametric uncertainty in this prediction.
For example, if both ${\cal C}$ and $h_C$ were 30\% smaller than the values we used,
the predicted rate would be four times smaller.
The general dependence of the leading-order branching ratio on  $|{\cal C} h_C|$ is shown in
Fig.~\ref{lodiff}(b).

\begin{figure}[tb]
\includegraphics[width=4.5in]{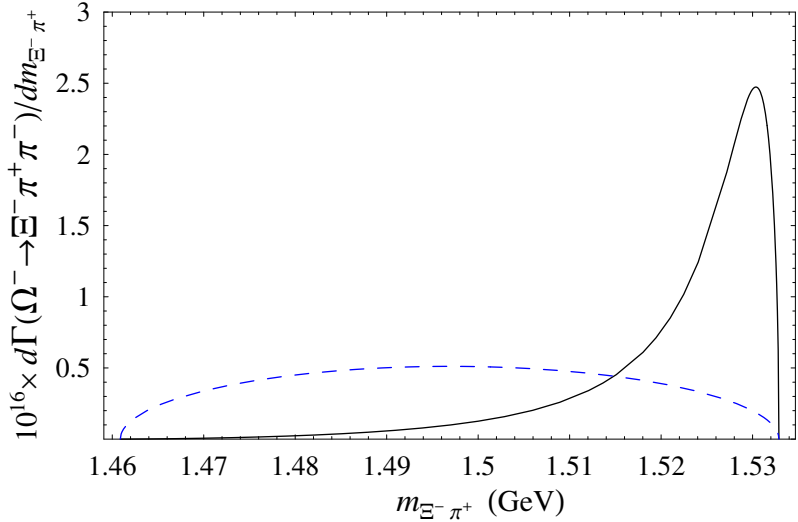}
\caption{\label{psdiff}Distributions of $\Xi^-\pi^+$ invariant-mass in
\,$\Omega^-\to\Xi^-\pi^+\pi^-$\, obtained from our leading-order amplitude (solid curve)
and from the assumption of uniform-phase-space decay distribution (dashed curve),
both normalized to yield \,${\cal B}(\Omega^-\to\Xi^-\pi^+\pi^-)=3.6\times 10^{-4}$\, .}
\end{figure}

The HyperCP data is not available in a format suitable for direct comparison with our
result due to detector effects.
However, their results indicate that a uniform phase-space distribution is a much
better fit to the data than a $\Xi^*$-dominated one~\cite{hypercp}.
In Fig.~\ref{psdiff} we plot the $m_{\Xi^-\pi^+}$ distributions resulting from our
leading-order amplitude (solid curve) and from assuming a uniform-phase-space decay
distribution (dashed curve), both normalized to reproduce the central value of HyperCP's result.
The structure of the leading-order amplitude, from Eq.~(\ref{loamps}), with all the terms
being proportional to ${\cal C}h_C$, is such that the $\Xi^*$ resonance is always the dominant
feature of the spectrum.
This leads us to investigate in the next section whether any of the next-to-leading-order
corrections can modify the predicted spectrum in the direction indicated by experiment.

\section{Calculation to next-to-leading order}

At next-to-leading order, ${\cal O}(p)$, there are two types of contributions.
The first type of contributions is that in which the weak transition occurs only between mesons.
To compute these contributions, we need the leading-order, ${\cal O}(p^2)$, strong and weak
Lagrangians for mesons, which are given respectively by~\cite{Gasser:1983yg,Cronin:1967jq}
\begin{subequations} \label{mesonl}
\begin{eqnarray}  \label{Ls'}
{\cal L}_{\rm s}' \,\,=\,\,
\mbox{$\frac{1}{4}$} f^2
\left\langle \partial^\mu\Sigma^\dagger\, \partial_\mu^{}\Sigma \right\rangle
+ \mbox{$\frac{1}{2}$} B_0^{} f^2 \left\langle M_+^{} \right\rangle  \,\,,
\end{eqnarray}
\begin{eqnarray} \label{weakcl1}
{\cal L}_{\rm w}' \,\,=\,\,
\gamma_8^{}f^2 \left\langle h\,\partial_\mu^{}\Sigma\,
\partial^\mu \Sigma^\dagger \right\rangle
\,\,+\,\,  {\rm H.c.}   \,\,,
\end{eqnarray}
\end{subequations}
where the parameter  $\gamma_8^{}$  is found from  $\,K\to\pi\pi$\, data to be
\begin{eqnarray}  \label{g8}
\gamma_8^{}  \,\,=\,\,  -7.8\times10^{-8}  \,\,,
\end{eqnarray}
the sign following from various predictions~\cite{g8}.

The contributions of the $\gamma_8^{}$ term are interesting  because the  \,$|\Delta S|=1$\,
weak transitions in the meson sector are larger than naive expectations.
In particular,  $\gamma_8^{}$ is several times larger than its naturally expected value
$\bigl(\sim$$1\times10^{-8}\bigr)$
and therefore could make its contributions numerically comparable to the lower-order ones.

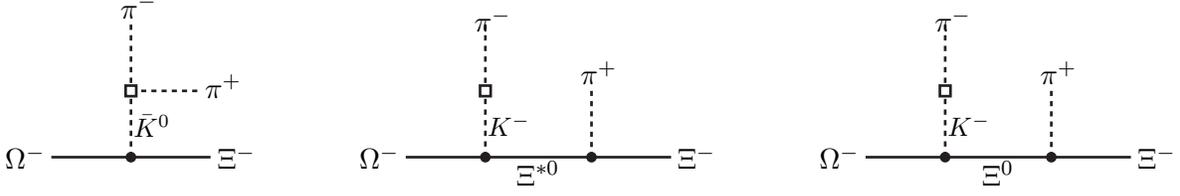
\begin{figure}[tb]
\begin{picture}(120,90)(-30,-10) \SetWidth{1}
\Text(-10,0)[]{\small$\Omega^-$}  \Line(0,0)(60,0) \Text(33,57)[]{\small$\pi^-$}
\DashLine(30,0)(30,50){2} \DashLine(30,25)(55,25){2} \Text(65,27)[]{\small$\pi^+$}
\Text(70,0)[]{\small$\Xi^-$}
\Text(38,12)[]{\footnotesize$\bar{K}^0$} \Vertex(30,0){2} \BBoxc(30,25)(4,4)
\end{picture}
\hfill
\begin{picture}(160,80)(-30,-10) \SetWidth{1}
\Text(-10,0)[]{\small$\Omega^-$} \Text(50,-6)[]{\small$\Xi^{*0}$}
\DashLine(30,0)(30,50){2} \Text(33,52)[]{\small$\pi^-$}
\DashLine(70,0)(70,25){2} \Text(73,32)[]{\small$\pi^+$}
\Line(0,0)(100,0) \Text(110,0)[]{\small$\Xi^-$} \Text(39,12)[]{\footnotesize$K^-$}
\Vertex(30,0){2} \Vertex(70,0){2}  \BBoxc(30,25)(4,4)
\end{picture}
\hfill
\begin{picture}(160,80)(-30,-10) \SetWidth{1}
\Text(-10,0)[]{\small$\Omega^-$} \Text(50,-6)[]{\small$\Xi^0$}
\DashLine(30,0)(30,50){2} \Text(33,52)[]{\small$\pi^-$}
\DashLine(70,0)(70,25){2} \Text(73,32)[]{\small$\pi^+$}
\Line(0,0)(100,0) \Text(110,0)[]{\small$\Xi^-$} \Text(39,12)[]{\footnotesize$K^-$}
\Vertex(30,0){2} \Vertex(70,0){2} \BBoxc(30,25)(4,4)
\end{picture}
\caption{\label{nlodiags}Diagrams contributing to  \,$\Omega^-\to\Xi^-\pi^+\pi^-$\,
at next-to-leading order in $\chi$PT.
Each solid dot represents a strong vertex from  ${\cal L}_{\rm s}^{}$  in Eq.~(\ref{Ls})
or ${\cal L}_{\rm s}'$  in  Eq.~(\ref{Ls'}), and
each square a weak vertex from  ${\cal L}_{\rm w}'$  in  Eq.~(\ref{weakcl1}).
}
\end{figure}

With weak vertices from the $\gamma_8^{}$ term alone, plus strong vertices from
${\cal L}_{\rm s}^{}$ and ${\cal L}_{\rm s}'$,  we derive the next-to-leading-order (NLO)
diagrams displayed in Fig.~\ref{nlodiags}.
They provide the NLO contributions to the $A_\pm$ and $B_\pm$ form factors in Eq.~(\ref{M_O}),
namely
\begin{subequations}  \label{nloamps}
\begin{eqnarray}
A_+^{(1)}  \,\,=\,\,
\frac{-{\cal C}\,\gamma_8^{}}{f^2}\,\, \frac{m_\pi^2-s_{+-}^{}}{m_K^2-s_{+-}^{}} \,\,,
\label{nloap}
\end{eqnarray}
\begin{eqnarray}
A_-^{(1)}  \,\,=\,\,  A_+^{(1)}  \,\,,
\label{nloam}
\end{eqnarray}
\begin{eqnarray}
B_+^{(1)}  \,\,=\,\,
\frac{-\cal C\,H}{3 f^2}\,\,
\frac{\gamma_8^{}\, m_\pi^2}
     {\bigl(m_K^2-m_\pi^2\bigr)\bigl(E_\Xi^{}+E_+^{}-\bar{m}_{\Xi^*}^{}\bigr)}  \,\,,
     \label{nlobp}
\end{eqnarray}
\begin{eqnarray} \label{nlobm}
B_-^{(1)}  &=&
\frac{-{\cal C}\, (D-F)}{f^2}\,\,
\frac{\gamma_8^{}\, m_\pi^2}
     {\bigl(m_K^2-m_\pi^2\bigr)\bigl(E_\Xi^{}+E_+^{}-m_\Xi^{}\bigr)}
\nonumber \\ && +\,\,
\frac{2\cal C\,H}{9 f^2}\,\,
\frac{\gamma_8^{}\, m_\pi^2}
     {\bigl(m_K^2-m_\pi^2\bigr)\bigl(E_\Xi^{}+E_+^{}-\bar{m}_{\Xi^*}^{}\bigr)}  \,\,.
\end{eqnarray}
\end{subequations}

There is another  type of NLO contribution to the amplitudes. It is given by diagrams similar
to those in Fig.~\ref{lodiags} in which one of the vertices is from a NLO Lagrangian.
Many of the parameters in NLO Lagrangians are not known, and so it is not possible at present
to include their contributions in a detailed way.
For example, the weak Lagrangian at ${\cal O}(p)$ that generates \,$\Omega^-\Xi^*\pi$\, and
\,$\Omega^-\Xi\pi$\, vertices is, as discussed in Appendix~\ref{Lw1},
\begin{eqnarray}  \label{tLw}
\tilde{\cal L}_{\rm w}'  &=&
\frac{h_{\Omega\Xi^*\pi}^{}}{f}\,v^\alpha\,\partial_\alpha^{}\pi^+\,\bar{\Xi}^{*0}\cdot\Omega^-
\,+\,
\frac{\tilde{h}_{\Omega\Xi^*\pi}^{}}{f}\, \partial_\alpha^{}\pi^+\,
\bar{\Xi}_\mu^{*0}\,2S_v^\alpha\, \Omega^{-\mu}
\nonumber \\ && +\,\,
\frac{h_{\Omega\Xi\pi}^{}}{f}\, \partial^\mu\pi^+\, \bar{\Xi}^0\,\Omega_\mu^-
\,\,+\,\, \cdots  \,\,,
\end{eqnarray}
where only the relevant terms are displayed and
$h_{\Omega\Xi^*\pi}$, $\tilde{h}_{\Omega\Xi^*\pi}$, and $h_{\Omega\Xi\pi}$  contain
unknown parameters.
The vertices occur in diagrams similar to the first one in Fig.~\ref{lodiags} with
intermediate  $\Xi^*$ and $\Xi$, yielding the NLO contributions
\begin{subequations} \label{nloextra}
\begin{eqnarray}
\tilde A_+^{(1)}  \,\,=\,\,
\frac{-{\cal C}\, h_{\Omega\Xi^*\pi}^{}\, E_-^{}}
     {\sqrt6\, f^2\,\bigl(E_\Xi^{}+E_+^{}-\bar{m}_{\Xi^*}^{}\bigr)}  \,\,,
\end{eqnarray}
\begin{eqnarray}
\tilde A_-^{(1)}  \,\,=\,\,  0  \,\,,
\end{eqnarray}
\begin{eqnarray}
\tilde B_+^{(1)}  \,\,=\,\,
\frac{-{\cal C}\,\tilde h_{\Omega \Xi^*\pi}}
     {\sqrt6\, f^2\, \bigl(E_\Xi^{}+E_+^{}-\bar{m}_{\Xi^{*}}\bigr)} \,\,,
\end{eqnarray}
\begin{eqnarray}  \label{wnloc}
\tilde B_-^{(1)}  \,\,=\,\,
\frac{(D-F)\,h_{\Omega\Xi\pi}^{}}{\sqrt 2\,f^2\, \bigl(E_\Xi^{}+E_+^{}-m_\Xi^{}\bigr)}
\,+\,
\frac{2{\cal C}\,\tilde h_{\Omega \Xi^*\pi}^{}}
     {3\sqrt6\, f^2\, \bigl(E_\Xi^{}+E_+^{}-\bar{m}_{\Xi^{*}}\bigr)} \,\,.
\end{eqnarray}
\end{subequations}
Numerically, we adopt the parametric variations
\begin{eqnarray} \label{hranges}
0  \,\,\le\,\,
\bigl|h_{\Omega\Xi^*\pi}^{}\bigr|,\, \bigl|\tilde h_{\Omega\Xi^*\pi}^{}\bigr|,\,
\bigl|h_{\Omega\Xi\pi}^{}\bigr|
\,\,\le\,\,  2\times 10^{-8}  \,\,,
\end{eqnarray}
where the upper limit is the expectation from naive dimensional analysis.

As mentioned above, there are additional NLO contributions that are not included in our
calculation because they depend on more unknown parameters.
We can still estimate the uncertainty in our results arising from those terms by allowing
the LO parameters to vary between their value as obtained from tree-level fits and their
value as obtained from one-loop fits.
For our numerics we will specifically consider parameter values
obtained from fits at one-loop order, which are available in the
literature~\cite{Jenkins:1991ne,Butler:1992pn,Egolf:1998vj}.
We begin by noticing that our results in Eqs.~(\ref{loamps}), (\ref{nloamps}),
and~(\ref{nloextra}) show that $f$ is a common factor affecting the overall normalization only.
Similarly, $\cal C$ is a common factor, except for the first term in Eq.~(\ref{wnloc}),
which is numerically small.
Consequently, we fix $f$ and $\cal C$ to their tree-level values, noting that the resulting
decay rate scales with an overall factor \,${\cal C}^2/f^4$.\,
In addition, we keep  $\gamma_8^{}$  at its value in Eq.~(\ref{g8}), as it is well determined.
Thus, the ranges of the strong parameters we consider are
\begin{eqnarray} \label{DFHranges}
0.21  \,\,\le\,\, D-F  \,\,\le\,\,  0.34 \,\,, \hspace{2em}
-2.4  \,\,\le\,\,  {\cal H}  \,\,\le\,\, -1.6 \,\,.
\end{eqnarray}
On the other hand, since the range of the weak parameter $h_C$ from one-loop fits is
large~\cite{Egolf:1998vj}, \,$-2\lesssim10^7\,h_C\lesssim4$,\,  we let it vary so as to
reproduce the experimental decay rates.

\begin{figure}[tb]
\begin{picture}(1,1)(0,0) \Text(10,190)[]{\footnotesize(a)} \end{picture}
\includegraphics[width=4.5in]{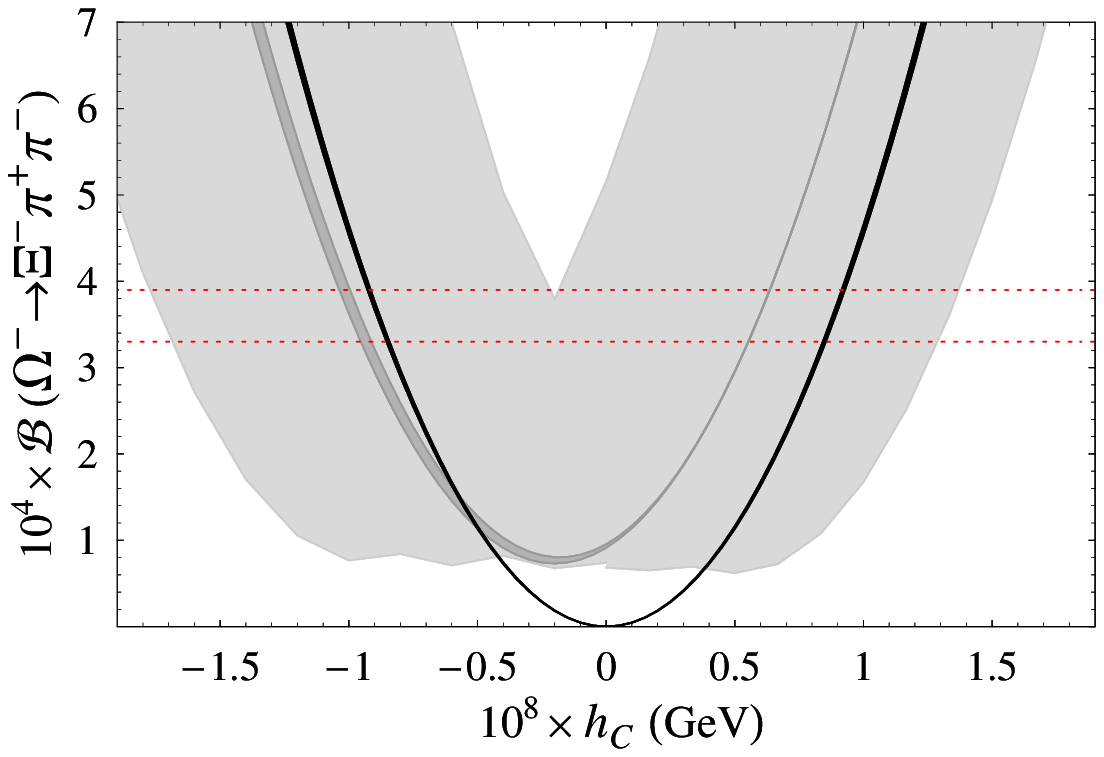}
\\ \bigskip
\begin{picture}(1,1)(0,0) \Text(75,130)[]{\footnotesize(b)~$h_C<0$} \end{picture}
\includegraphics[width=3.3in]{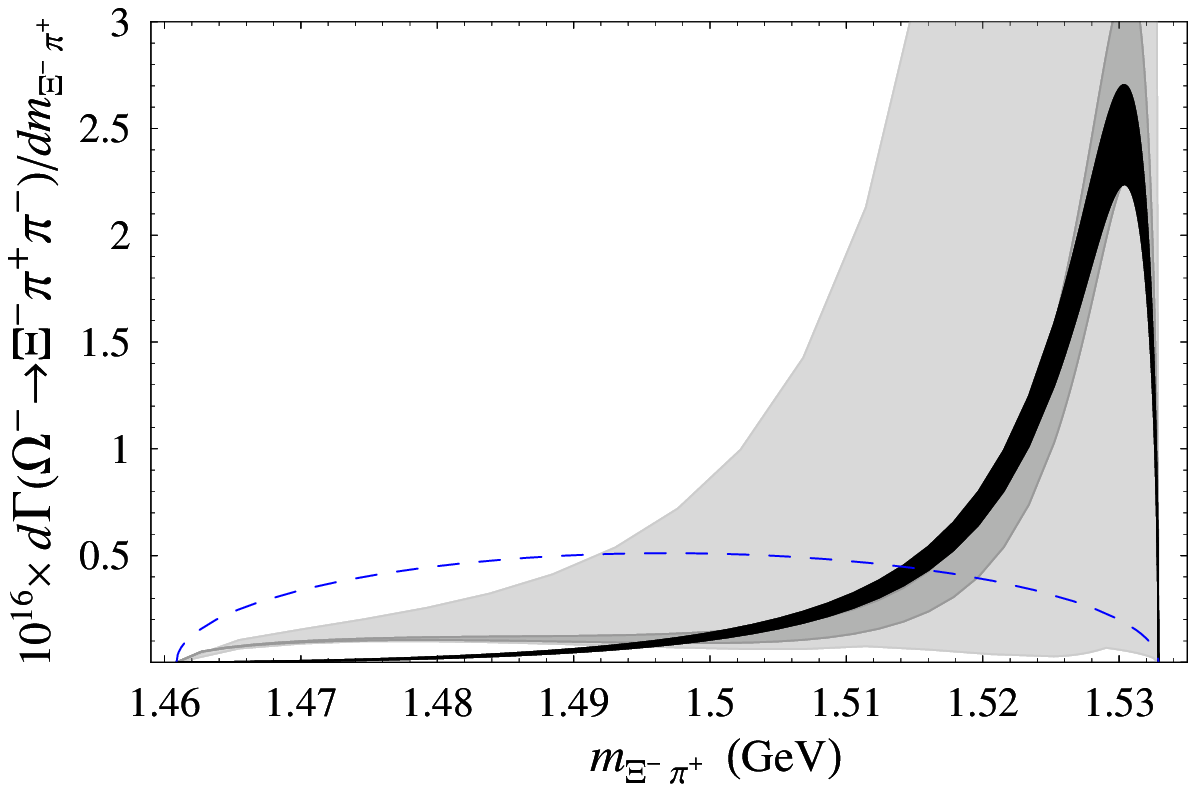}
\hspace*{-3em}
\begin{picture}(1,1)(0,0) \Text(67,130)[]{\footnotesize(c)~$h_C>0$} \end{picture}
\includegraphics[width=3.3in]{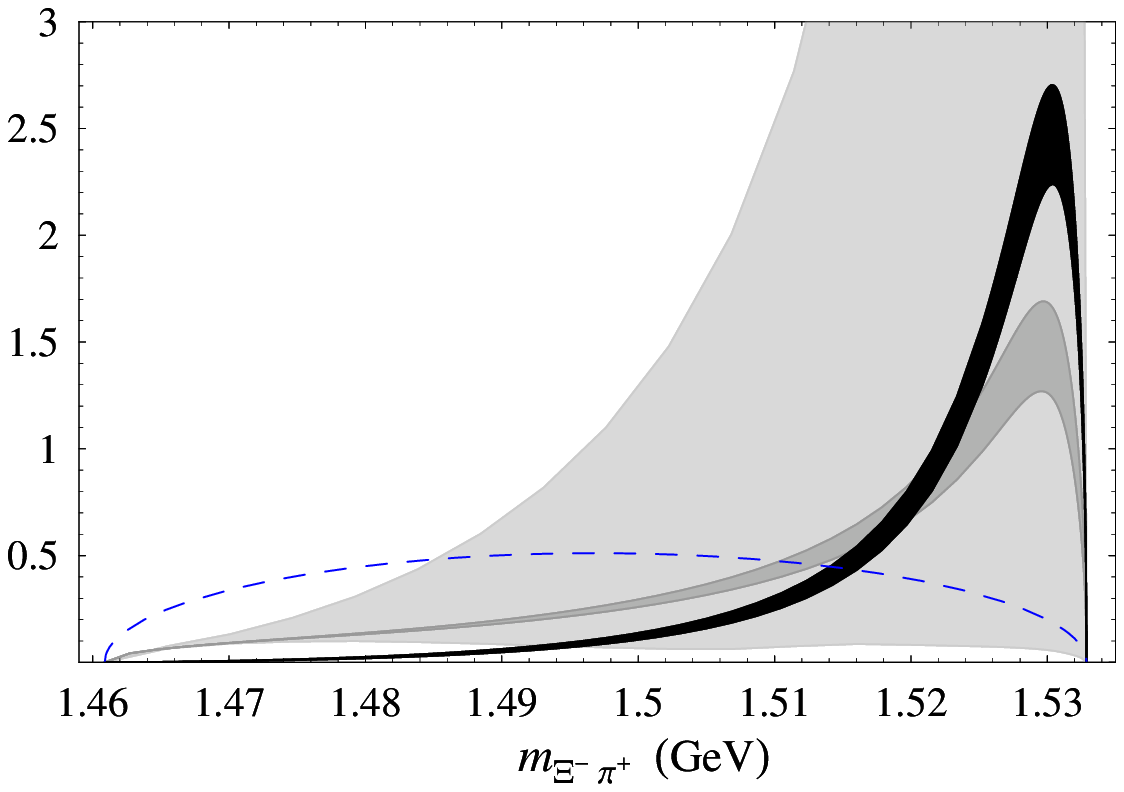}
\caption{\label{nlog8}(a) Branching ratios for  \,$\Omega^-\to\Xi^-\pi^+\pi^-$\,  and
(b,c) the corresponding distributions of $\Xi^-\pi^+$ invariant-mass.
The black (dark gray) bands come from the LO amplitude only (the LO amplitude and
the $\gamma_8^{}$ terms in the NLO amplitude), and the light-gray bands result from
the LO and NLO amplitudes we consider, as described in the text.
The dotted lines in (a) bound the range implied by the preliminary HyperCP data.
The dashed curves in (b) and (c) have been reproduced from Fig.~\ref{psdiff}.}
\end{figure}

In Fig.~\ref{nlog8}(a) we display the branching ratios calculated from the leading-order (LO)
and NLO amplitudes above.
The black (dark gray) band in the figure shows the effects of the parametric variations
given in Eq.~(\ref{DFHranges}) on the branching ratio obtained from the LO amplitude alone
(the LO amplitude and only the $\gamma_8^{}$ terms in the NLO amplitude).
The light-gray region results from the LO and NLO amplitudes considered above and
varying the parameters according to Eqs.~(\ref{hranges}) and~(\ref{DFHranges}).
The dotted lines in this figure bound the range
\,$3.3\le10^4\,{\cal B}(\Omega^-\to\Xi^-\pi^+\pi^-)\le3.9$\,
implied by the preliminary HyperCP data.
Evidently, this data can be reproduced in the three cases.

The corresponding $m_{\Xi^-\pi^+}$ distributions are plotted in Figs.~\ref{nlog8}(b) and~(c)
for \,$h_C<0$\,  and \,$h_C>0$,\, respectively, with the variations of the other parameters
for the different bands being the same as in Fig.~\ref{nlog8}(a).
The $h_C$ ranges used in (b) and~(c) are \,$0.84<10^8\,|h_C|<0.92$\, for the black bands,
\,$-1.05<10^8\,h_C<-0.90$\, and \,$0.55<10^8\,h_C<0.65$\, for the dark-gray bands, and
\,$-1.8<10^8\,h_C<0$\, and \,$0<10^8\,h_C<1.4$\, for the light-gray bands, all of which have
been inferred from the corresponding bands in~(a).
The figures indicate that some softening of the $\Xi^*$ dominance in the spectrum is possible
with the inclusion of higher-order contributions.

\section{Conclusions}

We have evaluated the decay  \,$\Omega^-\to\Xi^-\pi^+\pi^-$\,  in heavy-baryon chiral
perturbation theory. At leading order, we found a~spectrum dominated by the $\Xi^*(1530)$,
as had been suggested before. This shape is in conflict with the recent preliminary data
from HyperCP. The total branching ratio is also in conflict with experiment for the central
values of ${\cal C}$ and $h_C$, but it suffers from a large parametric uncertainty.
This uncertainty, however, does not affect the shape of the $m_{\Xi^-\pi^+}$ invariant mass
distribution.

A complete calculation at next-to-leading-order contains too many unknown parameters to be
phenomenologically useful. We have investigated the effect of the NLO corrections in three
different ways.
First, we considered the diagrams in which the weak transition occurs in the meson sector.
These corrections
are induced by the low-energy constant $\gamma_8^{}$ which is  known from kaon decay.
Second, we considered the NLO terms in the weak chiral Lagrangian which introduce three
new effective constants. We studied the effect of these constants by varying their value
between zero and the value suggested by naive dimensional analysis.
Third and last, we varied the LO parameters in ranges that included their values as
determined from tree-level and one-loop fits to other hyperon decay modes.
The difference between the two kinds of fit is indicative of the size of NLO counterterms
that we have not included explicitly.
When all these factors are considered, we have found that it is possible to lower
the branching ratio and soften the importance of the $\Xi^*$ in the $m_{\Xi^-\pi^+}$
distribution, as suggested by the data.
Beyond this, we can only encourage the HyperCP collaboration to fit their data
to our result, given in Eqs.~(\ref{loamps}), (\ref{nloamps}), and~(\ref{nloextra}).

\begin{acknowledgments}

The work of O.A. and G.V. was supported in part by DOE under contract number DE-FG02-01ER41155.
We thank D.~Atwood, O.~Kamaev, D.~Kaplan, and S.~Prell for useful conversations.

\end{acknowledgments}

\appendix

\section{Derivation of next-to-leading-order weak Lagrangian\label{Lw1}}

The NLO weak Lagrangian generating the weak  \,$\Omega^-T\varphi$\,  vertices generally
contains the Dirac structures  \,$\bar T^\mu\, v\cdot{\cal A}\,T_\mu$\,  and
\,$\bar T^\mu\, 2S\cdot{\cal A}\,T_\mu$.\,
The only possible SU(3) building-blocks needed to construct it are therefore the tensors
$\bar{T}_{abc}$, ${\cal A}_{de}$,  and $T_{fgh}$.
Employing standard techniques~\cite{tdlee}, we treat the combination
\,$\bar T_{abc}\, {\cal A}_{de}\, T_{fgh}$\, as a tensor product
\,$\bigl(\overline{10}\otimes8\bigr)\otimes10$.\,
Thus we find four different operators that transform as octets, whose irreducible
representations are
\begin{eqnarray}
\begin{array}{c} \displaystyle
\bigl({\cal O}_1^{}\bigr)_{ab} \,\,=\,\,
\epsilon_{bmn}^{}\, \bar o_{m,r}^{}\, T_{anr}^{} \,\,,
\hspace{2em}
\bigl({\cal O}_2^{}\bigr)_{ab} \,\,=\,\,
\bar d_{bmn}^{}\, T_{amn}^{}
- \mbox{$\frac{1}{3}$}\, \delta_{ab}^{}\, \bar d_{mno}^{}\, T_{mno}^{} \,\,,
\vspace{2ex} \\   \displaystyle
\bigl({\cal O}_3^{}\bigr)_{ab} \,\,=\,\,
\epsilon_{bmn}^{}\, \bar\tau_{am,op}^{}\, T_{nop}^{} \,\,,
\hspace{2em}
\bigl({\cal O}_4^{}\bigr)_{ab} \,\,=\,\, \bar\theta_{a,bmno}^{}\, T_{mno}^{}  \,\,,
\end{array}
\end{eqnarray}
where
\begin{eqnarray}
\bar o_{a,b}^{} \,\,=\,\, \epsilon_{amn}^{}\, \bar{T}_{bmo}^{}\, {\cal A}_{on}^{} \,\,,
\hspace{3em}
\bar d_{abc}^{} \,\,=\,\,
\bar{T}_{abm}^{}\, {\cal A}_{mc}^{}+\bar{T}_{acm}^{}\, {\cal A}_{mb}^{}
+ \bar{T}_{bcm}^{}\, {\cal A}_{ma}^{}  \;,
\end{eqnarray}
\begin{eqnarray}
\bar\tau_{ab,cd}^{} &=&
\bar{T}_{cdm}^{}\,\bigl(\epsilon_{amo}^{}\,{\cal A}_{bo}^{}
                        + \epsilon_{bmo}^{}\,{\cal A}_{ao}^{}\bigr)
\nonumber \\ && -\,\,
\mbox{$\frac{1}{5}$} \bigl( \delta_{ac}^{} \bar{T}_{dmn}^{}\, \epsilon_{bmo}^{}
                   + \delta_{ad}^{} \bar{T}_{cmn}^{}\, \epsilon_{bmo}^{}
                   + \delta_{bc}^{} \bar{T}_{dmn}^{}\, \epsilon_{amo}^{}
        + \delta_{bd}^{} \bar{T}_{cmn}^{}\, \epsilon_{amo}^{} \bigr)\, {\cal A}_{no}^{} \,\,,
\end{eqnarray}
\begin{eqnarray}
\bar\theta_{a,bcde}^{} &=&
\bar{T}_{bcd}^{}\, {\cal A}_{ae}^{} + \bar{T}_{bce}^{}\, {\cal A}_{ad}^{}
+ \bar{T}_{bde}^{}\, {\cal A}_{ac}^{} + \bar{T}_{cde}^{}\, {\cal A}_{ab}^{}
\nonumber \\ && -\,\,
\mbox{$\frac{1}{6}$}\, \delta_{ab}^{}
\bigl( \bar{T}_{cdm}^{}\, {\cal A}_{me}^{} + \bar{T}_{cem}^{}\, {\cal A}_{md}^{}
+ \bar{T}_{dem}^{}\, {\cal A}_{mc}^{} \bigr)
\nonumber \\ && -\,\,
\mbox{$\frac{1}{6}$}\, \delta_{ac}^{}
\bigl( \bar{T}_{bdm}^{}\, {\cal A}_{me}^{} + \bar{T}_{bem}^{}\, {\cal A}_{md}^{}
+ \bar{T}_{dem}^{}\, {\cal A}_{mb}^{} \bigr)
\nonumber \\ && -\,\,
\mbox{$\frac{1}{6}$}\, \delta_{ad}^{}
\bigl( \bar{T}_{bcm}^{}\, {\cal A}_{me}^{} + \bar{T}_{cem}^{}\, {\cal A}_{mb}^{}
+ \bar{T}_{bem}^{}\, {\cal A}_{mc}^{} \bigr)
\nonumber \\ && -\,\,
\mbox{$\frac{1}{6}$}\, \delta_{ae}^{}
\bigl( \bar{T}_{bcm}^{}\, {\cal A}_{md}^{} + \bar{T}_{cdm}^{}\, {\cal A}_{mb}^{}
+ \bar{T}_{bdm}^{}\, {\cal A}_{mc}^{} \bigr)  \;.
\end{eqnarray}
The tensors  $({\cal O}_{1,2,3,4})_{ab}$ and  $\bar o_{ab}$
are all traceless,  $\bar d_{abc}$ is fully symmetric in its indices,
$\bar\tau_{ab,cd}$  satisfies the symmetry relation
\,$\bar\tau_{ab,cd}=\bar\tau_{ba,cd}=\bar\tau_{ab,dc}=\bar\tau_{ba,dc}$\,
and tracelessness condition  \,$\bar\tau_{ab,cb}=0$,\,
and   $\bar\theta_{a,bcde}$  is symmetric in its $bcde$ indices
and satisfies  \,$\bar\theta_{a,abcd}=0$.\,

The only possible building blocks needed to construct the NLO weak Lagrangian
generating the weak  \,$\Omega^-B\varphi$\,  vertices are the tensors $\;\bar{B}_{ab}$,
${\cal A}_{cd}$, and $T_{def}$.
Treating the combination  \,$\bar B_{ab}\, {\cal A}_{cd}\, T_{def}$\, as a tensor
product \,$(8\otimes8)\otimes10$,\, we find four different operators that transform as octets,
whose irreducible representations are
\begin{eqnarray}
\begin{array}{c} \displaystyle
\bigl({\cal O}_1'\bigr)_{ab} \,\,=\,\, \epsilon_{bmn}^{}\, \bar{\cal D}_{mo}^{}\, T_{ano}^{} \,\,,
\hspace{2em}
\bigl({\cal O}_2'\bigr)_{ab} \,\,=\,\, \epsilon_{bmn}^{}\, \bar{\cal F}_{mo}^{}\, T_{ano}^{} \,\,,
\vspace{2ex} \\   \displaystyle
\bigl({\cal O}_3'\bigr)_{ab} \,\,=\,\,
\bar t_{bmn}^{}\, T_{amn}^{}-\mbox{$\frac{1}{3}$}\, \delta_{ab}^{}\, \bar t_{mno}^{}\, T_{mno}^{} \,\,,
\hspace{2em}
\bigl({\cal O}_4'\bigr)_{ab}  \,\,=\,\,
\epsilon_{bmn}^{}\, \bar{\cal T}_{am,op}^{}\, T_{nop}^{} \,\,,
\end{array}
\end{eqnarray}
where
\begin{eqnarray}
\bar{\cal D}_{ab}^{} \;=\;  \bigl\{\bar{B}, {\cal A} \bigr\}_{ab}
      - \mbox{$\frac{2}{3}$}\, \bigl\langle\bar{B}{\cal A}\bigr\rangle\, \delta_{ab}^{} \,\,,
\hspace{3em}
\bar{\cal F}_{ab}^{} \;=\;  \bigl[\bar{B}, {\cal A} \bigr]_{ab}   \;,
\end{eqnarray}
\begin{eqnarray}
\bar t_{abc}^{} &=&
\epsilon_{amn}^{} \left( \bar{B}_{mb}^{} {\cal A}_{nc}^{}
                        + \bar{B}_{mc}^{} {\cal A}_{nb}^{} \right)
+ \epsilon_{bmn}^{} \left( \bar{B}_{mc}^{} {\cal A}_{na}^{}
                          + \bar{B}_{ma}^{} {\cal A}_{nc}^{} \right)
\nonumber \\ && +\,\,
\epsilon_{cmn}^{} \left( \bar{B}_{ma}^{} {\cal A}_{nb}^{}
                        + \bar{B}_{mb}^{} {\cal A}_{na}^{} \right)   \;,
\end{eqnarray}
\begin{eqnarray}
\bar{\cal T}_{ab,cd}^{} &=&
\bar{B}_{ac}^{} {\cal A}_{bd}^{} + \bar{B}_{ad}^{} {\cal A}_{bc}^{}
+ \bar{B}_{bc}^{} {\cal A}_{ad}^{} + \bar{B}_{bd}^{} {\cal A}_{ac}^{}
\nonumber \\ && -\,\,
\mbox{$\frac{1}{5}$} \left(
\delta_{ac}^{} \bar{\cal D}_{bd}^{}
+ \delta_{ad}^{} \bar{\cal D}_{bc}^{}
+ \delta_{bc}^{} \bar{\cal D}_{ad}^{}
+ \delta_{bd}^{} \bar{\cal D}_{ac}^{}
\right)
\,-\,
\mbox{$\frac{1}{6}$} \left( \delta_{ac}^{}\delta_{bd}^{}
                   + \delta_{ad}^{}\delta_{bc}^{} \right)
\bigl\langle\bar{B}{\cal A}\bigr\rangle  \,\,.
\end{eqnarray}
The tensors  $({\cal O}_{1,2,3,4}')_{ab}$,  $\bar{\cal D}_{ab}$, and  $\bar{\cal F}_{ab}$
are all traceless,  $\bar t_{abc}$ is fully symmetric in its indices, and
$\bar{\cal T}_{ab,cd}$  satisfy the symmetry relation
\,$\bar{\cal T}_{ab,cd}=\bar{\cal T}_{ba,cd}=\bar{\cal T}_{ab,dc}=\bar{\cal T}_{ba,dc}$\,
and tracelessness condition  \,$\bar{\cal T}_{ab,cb}=0$.\,

The resulting NLO weak Lagrangian that contributes to  \,$\Omega^-\to\Xi^*\pi,\Xi\pi$\,  and
transforms as $(8_{\rm L},1_{\rm R})$  is then
\begin{eqnarray}
\tilde{\cal L}_{\rm w}'  &=&
\left\langle \xi^\dagger h\xi \left( h_1^{}\, {\cal O}_1^{} + h_2^{}\, {\cal O}_2^{}
+ h_3^{}\, {\cal O}_3^{} + h_4^{}\, {\cal O}_4^{} \right) \right\rangle
\nonumber \\ && +\,\,
\left\langle \xi^\dagger h\xi \left( \tilde h_1^{}\, \tilde{\cal O}_1^{}
+ \tilde h_2^{}\, \tilde{\cal O}_2^{} + \tilde h_3^{}\, \tilde{\cal O}_3^{}
+ \tilde h_4^{}\, \tilde{\cal O}_4^{} \right) \right\rangle
\nonumber \\ && +\,\,
\left\langle \xi^\dagger h\xi \left( h_1'\, {\cal O}_1' + h_2'\, {\cal O}_2'
+ h_3'\, {\cal O}_3' + h_4'\, {\cal O}_4' \right) \right\rangle  \,\,,
\end{eqnarray}
where  ${\cal O}_i$  and  $\tilde{\cal O}_i$  contain the Dirac structures
\,$\bar T^\mu\,v\cdot{\cal A}\,T_\mu$\,  and  \,$\bar T^\mu\,2S\cdot{\cal A}\,T_\mu$,
respectively, and  $h_i^{}$, $\tilde h_i^{}$, and  $h_i'$  and   are free parameters.
Expanding the Lagrangian yields
\begin{eqnarray}
\tilde{\cal L}_{\rm w}'  &=&
\frac{h_{\Omega\Xi^*\pi}^{}}{f}\,v^\alpha\,\partial_\alpha^{}\pi^+\,
\bar{\Xi}^{*0}\cdot\Omega^-
\,+\,
\frac{\tilde{h}_{\Omega\Xi^*\pi}^{}}{f}\, \partial_\alpha^{}\pi^+\,
\bar{\Xi}_\mu^{*0}\,2S_v^\alpha\, \Omega^{-\mu}
\nonumber \\ && +\,\,
\frac{h_{\Omega\Xi\pi}^{}}{f}\, \partial^\mu\pi^+\, \bar{\Xi}^0\,\Omega_\mu^-
\,\,+\,\, \cdots  \,\,,
\end{eqnarray}
where
\begin{eqnarray}
\begin{array}{c} \displaystyle
\stackrel{\scriptscriptstyle(\sim)}{h}_{\Omega\Xi^*\pi}^{}  \,\,=\,\,
\frac{\stackrel{\scriptscriptstyle(\sim)}{h}_1^{}}{\sqrt6}
+ \frac{\stackrel{\scriptscriptstyle(\sim)}{h}_2^{}}{\sqrt6}
- \frac{7\,\stackrel{\scriptscriptstyle(\sim)}{h}_3^{}}{5\sqrt6}
- \frac{\stackrel{\scriptscriptstyle(\sim)}{h}_4^{}}{2\sqrt6}  \,\,,
\vspace{2ex} \\   \displaystyle
h_{\Omega\Xi\pi}^{}  \,\,=\,\,
\frac{-h_1'}{\sqrt2}+\frac{h_2'}{\sqrt2}-\sqrt2\, h_3'+\frac{\sqrt2\,h_4'}{5}  \,\,.
\end{array}
\end{eqnarray}

\end{document}